\documentclass[journal]{IEEEtran}
\usepackage{amsfonts, datetime, mathtools, graphicx, amsmath, amsthm, amssymb, bm, amsbsy, breqn, accents, float, listings, courier, lscape, multirow, multicol, longtable, titling, dsfont, bbold, cite, soul, color, tikz, scalerel,empheq, paralist}
\usetikzlibrary{shapes, arrows}
\allowdisplaybreaks
\newcommand{\R}{\mathbb{R}}

\newcommand{\e}{\mathrm{e}}

\newcommand{\scrJ}{\mathcal{J}}
\newcommand{\scrL}{\mathcal{L}}

\newcommand{\scrN}{\mathcal{N}}
\newcommand{\scrO}{\mathcal{O}}

\newcommand{\boldR}{\mathbf{R}}

\newcommand{\bracketRound}[1]{\left(#1\right)}
\newcommand{\bracketSquare}[1]{\left[#1\right]}

\newcommand{\Expect}[1]{\mathbb{E}{\bracketSquare{#1}}}
\newcommand{\Prob}{\mathbb{P}}

\newcommand{\probMatrix}{P}
\newcommand{\jammerStrategy}{J}
\newcommand{\jammerStrategyOption}{j}
\newcommand{\jammerStrategyOptionSet}{\scrJ}
\newcommand{\radarStrategy}{\pi}
\newcommand{\radarObservation}{R}
\newcommand{\boldRadarObservation}{\boldR}

\newcommand{\radarUtility}{\phi}
\newcommand{\jammerUtility}{\psi}

\newcommand{\Variance}{\Sigma}

\newcommand{\maxEV}[1]{\lambda\bracketRound{#1}}
\newcommand{\indn}{}
\newcommand{\desired}{n}

\newcommand{\SNRn}{\overline{\operatorname{SNR}}^{(\timeIndexTwo)}}
\newcommand{\condPMF}[2]{\displaystyle{\Prob(\boldRadarObservation=#1\mid\textbf{\jammerStrategy}=#2)}}
\newcommand{\condPMFhat}[2]{\displaystyle{\hat{\Prob}(\boldRadarObservation=#1\mid\textbf{\jammerStrategy}=#2)}}

\newcommand{\radarStrategyRandomX}{\operatorname{e}_{\boldRadarObservation}^\prime x}
\newcommand{\radarStrategyRandom}{\operatorname{e}_{\boldRadarObservation}^\prime \radarStrategy}

\newcommand{\timeIndexOne}{k}
\newcommand{\timeIndexTwo}{n}
\newcommand{\timeIndexThree}{t}
\newcommand{\eqrefPAgeneral}{\eqref{eq:pa-general}}
\newcommand{\eqrefPAscSIMPLIFIED}{\eqref{eq:pa-sc-simplified}}
\newcommand{\eqUTILITYfunctions}{\eqref{eq:utility-functions}}

\newtheorem{theorem}{Theorem}

\newtheorem{lemma}{Lemma}
\newtheorem{definition}{Definition}

\newcommand*\widefbox[1]{\fbox{\hspace{2em}#1\hspace{2em}}}
\date{}

\begin{document}
	
	\title{Principal Agent Problem as a Principled Approach to Electronic Counter-Countermeasures in Radar}
	
	\author{Anurag~Gupta\thanks{Anurag Gupta and Vikram Krishnamurthy are with the School of Electrical \& Computer Engineering, Cornell University, Ithaca NY, 14853, USA.  (e-mail: ag2589@cornell.edu; vikramk@cornell.edu).},
	and Vikram~Krishnamurthy\footnotemark[1],~\IEEEmembership{Fellow,~IEEE}}
	
	\maketitle	
	\begin{abstract}
		Electronic countermeasures (ECM) against a radar are actions taken by an adversarial jammer to mitigate the effective utilization of the electromagnetic spectrum by the radar. On the other hand, electronic counter-countermeasures (ECCM) are actions taken by the radar to mitigate the impact of electronic countermeasures (ECM) so that the radar can continue to operate effectively. The main idea of this paper is to show that ECCM involving a radar and a jammer can be formulated as a  principal-agent problem (PAP) - a problem widely studied in microeconomics. With the radar as the principal and the jammer as the agent, we design a PAP to optimize the radar's ECCM strategy in the presence of a jammer. The radar seeks to optimally trade-off signal-to-noise ratio (SNR) of the target measurement with the measurement cost: cost for generating radiation power for the pulse to probe the target.
		We show that for a suitable choice of utility functions, PAP is a convex optimization problem.
		Further, we analyze the structure of the PAP and provide sufficient conditions under which the optimal solution is an increasing function of the jamming power observed by the radar; this enables computation of the radar's optimal ECCM within the class of increasing affine functions at a low computation cost. Finally, we illustrate the PAP formulation of the radar's ECCM problem via numerical simulations. We also use simulations to study a radar's ECCM problem wherein the radar and the jammer have mismatched information.
	\end{abstract}
	
	
	\IEEEpeerreviewmaketitle
	
\section{Introduction}
Since radars operate in a shared electromagnetic environment, they are susceptible to \textit{electronic countermeasures} (ECM): actions taken by an adversarial jammer to prevent effective utilization of the electromagnetic spectrum and thereby decrease the measurement accuracy of the radar.
	Hence, modern radars are often equipped with {\em electronic counter-countermeasures} (ECCM): strategies to mitigate the impact of ECM by an adversarial jammer.
	A list of standard ECM and ECCM techniques are summarized in \cite{1978:SJ} and \cite{2008:MS}. 
	
	\textit{Why Principal-Agent Problem (PAP)?}
	Our main idea is to formulate the radar's ECCM problem as a PAP. The PAP \cite{1995:AM} has been studied extensively in micro-economics to enforce a  contract between two entities in  labor contracts \cite{2005:GM}, insurance market \cite{2003:MH}, and  differential privacy \cite{2014:CD-AR}. At the core of the PAP lies information asymmetry: the principal only views the agent's action in noise, so the principal needs to write a contract with suitable incentives to induce action from the agent that would maximize its utility.
	
	In this paper, we model the radar as the principal and the adversarial jammer as the agent. The radar is interested in tracking a target of interest and maximizing its measurement accuracy. The jammer injects jamming power (ECM) into the environment to decrease the measurement accuracy of the radar. The radar, in turn, observes the action of the jammer in noise and takes action (ECCM) to counter the ECM of the jammer. The radar should ensure an optimal balance between measurement accuracy and \textit{measurement cost}: cost to generate radiation power for the pulse to probe the target. It is important that we incorporate measurement cost while designing ECCM as a large number of measurements have to be made for continuous monitoring of targets. Similarly, the jammer has to consider a \textit{jamming cost} for generating radiation noise power. Again it is important to include it in the jammer's utility function as the jamming has to be done continuously to reduce the radar's tracking accuracy.
	
	The information asymmetry between the radar and the jammer motivates  PAP as a way to study the radar's ECCM problem. The PAP constitutes a  principled approach to  ECCM: the PAP formulation captures the information asymmetries in ECCM, it yields a formulation of optimal ECCM problem as a convex optimization problem, and the resulting solution has a useful stochastic dominance structure that can be exploited for computing a constrained solution at a low computation cost. The PAP is also flexible to accommodate additional constraints on the information of the radar and the jammer: we study through numerical examples a radar's ECCM problem when the radar and the jammer have mismatched information.
	
	\subsection*{Related Work}
	Regarding  ECCM  for radars,
	\cite{2012:XS-et-al}, \cite{2015:HG-et-al} analyze the power allocation problem for a MIMO radar and a jammer as a game both for complete and incomplete information. \cite{2016:AD-et-al} generalizes the setting for a network of radars and jammers. A game based on a two-stage optimization method was considered in \cite{2019:KL-et-al} with either the radar or the jammer as the leader.
	
	Related to pulse-level implementation aspects of the radar's ECCM problem, an ECCM scheme based on time-frequency analysis was proposed in \cite{2014:SG-et-al}  for a particular type of deceptive jamming. Radar waveform design to combat barrage jamming was examined in \cite{2019:SC-et-al}, \cite{2018:FB-et-al}.
	
	To the best of our knowledge, the PAP approach to model the adversarial interaction between a radar's ECCM strategy and a jammer's ECM has not been explored in literature. The PAP framework yields a tractable convex optimization problem and also allows us to analyze the structure of the optimal ECCM. Moreover, it overcomes the non-uniqueness of the solution faced with the game-theoretic approach without making strong assumptions on the utility functions.
	Hence the motivation for this paper.
	
	\subsection*{Organization and Main Results}
	Sec.\ref{sec:radar_jammer_ew} describes the PAP for the radar's ECCM problem in the presence of a jammer performing ECM. Using a suitable choice of utility functions, we derive an equivalent convex optimization problem for the PAP. This allows us to derive analytical results and numerically solve the radar's optimal ECCM strategy using convex optimization solvers.
	
	In Sec.\ref{sec:structural-results},	we exploit the structure of the PAP  to characterize the radar's optimal ECCM strategy: the radar's optimal ECCM strategy is an increasing function of the jamming power observed by the radar. This enables us to parametrize the solution and find the radar's optimal ECCM strategy within a constrained class of functions at a low computation cost. 
	
	Finally, Sec.\ref{sec:simulation} illustrates the PAP  model for the ECCM problem using numerical examples.  We apply the structural result presented in  Sec.\ref{sec:structural-results} to compute a constrained solution to the radar's ECCM strategy. We also study a radar's ECCM problem via numerical examples wherein the radar and the jammer have mismatched information.
\section{Electronic Counter-Countermeasure Model}
\label{sec:radar_jammer_ew}
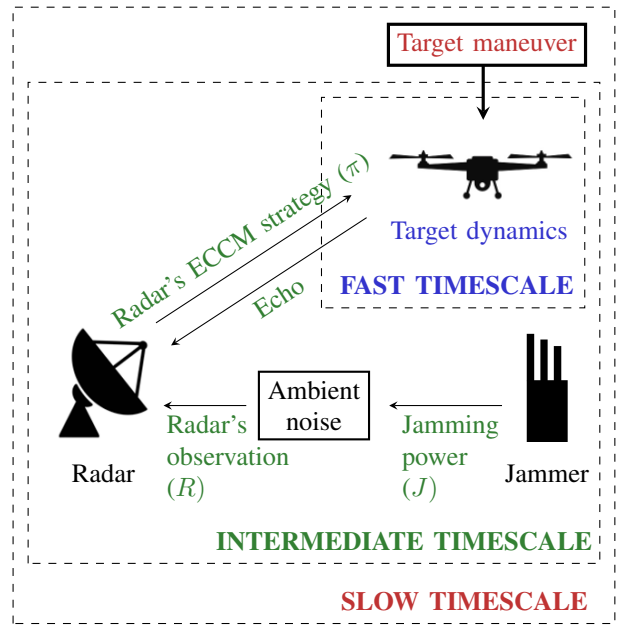
\begin{figure}
	\begin{center}
		\begin{tikzpicture}
			\tikzstyle{arrow} = [->,>=stealth];
			\definecolor{intermediate}{rgb}{0.2,0.5,0.2};
			\definecolor{fast}{rgb}{0.2,0.2,0.8};
			\definecolor{slow}{rgb}{0.75,0.2,0.2};
			\draw [dashed] (0,-0.2) rectangle(8,8);
			\draw [dashed] (0.2,0.6) rectangle(7.8,7);
			\draw [dashed] (4.1,4) rectangle(7.6,6.8);
			\node[inner sep=0] (target) at (6.2,5.8)
			{\includegraphics[width=0.15\textwidth]{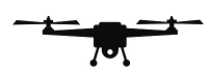}};
			\node[draw, align=right, draw=none, color=fast] at (6.2,5) {Target dynamics};
			\node[inner sep=0] (radar) at (1.2,3)
			{\includegraphics[width=0.1\textwidth]{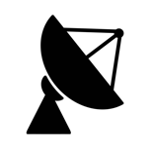}};
			\node[draw, align=right, draw=none] at (1.2,1.8) {Radar};
			\node[inner sep=0] (jammer) at (7.1,2.94)
			{\includegraphics[width=0.03\textwidth]{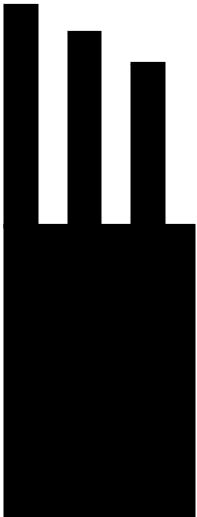}};
			\node[draw, align=right, draw=none] at (7.1,1.8) {Jammer};
			\draw [arrow] (1.9,3.8) -- node[anchor=south, sloped, color=intermediate] {Radar's ECCM strategy ($\pi$)} (4.5,5.5);
			\draw [arrow] (4.7,5.2) -- node[anchor=north, sloped, color=intermediate] {Echo} (2.1,3.5);
			\draw [arrow] (6.5,2.7) -- node[anchor=north, sloped, color=intermediate, align=right, xshift=12] {\parbox{2cm}{Jamming\\power\\($\jammerStrategy)$}} (5,2.7);
			\draw [arrow] (3.1,2.7) -- node[anchor=north, sloped, color=intermediate, xshift=14] {\parbox{2cm}{Radar's\\observation\\($\radarObservation)$}} (2,2.7);
			\node[draw, align=center, thick] at (4,2.7) {Ambient\\noise};
			\node[draw, align=right, draw=none, color=fast, font=\bfseries] at (5.9,4.3) {FAST TIMESCALE};
			\node[draw, align=right, draw=none, color=intermediate, font=\bfseries] at (5.2,0.9) {INTERMEDIATE TIMESCALE};
			\node[draw, align=right, draw=none, color=slow, font=\bfseries] at (6,0.1) {SLOW TIMESCALE};
			\node[draw, align=right, draw=black, very thick, text=slow] at (6.3,7.5) {Target maneuver};
			\draw [-stealth, very thick] (6.25,7.24) -- (6.25,6.5);
		\end{tikzpicture}
	\end{center}
	\caption{Three timescale radar-jammer problem for tracking a target. Target dynamics/transient occurs on the fast timescale; radar-jammer PAP is solved at the intermediate timescale; target is maneuvered at the slow timescale.}
	\label{fig:block_diagram}
\end{figure}

In this section, we formulate the PAP  for optimizing the  radar's ECCM 
strategy. As shown in Fig.~\ref{fig:block_diagram}, there are three independent entities in our problem formulation:
\begin{enumerate}
\item
a radar \item a jammer 
\item and a target (or multiple targets)
\end{enumerate}
In the formulation below, the radar and jammer interact while the target evolves independently. The main idea is that the radar exploits this interaction to mitigate the effect of ECM by using ECCM; it is this ECCM  aspect that we model as a PAP below.
For simplicity, we describe our model for a single target; Sec.\ref{sec:multiple-targets} describes generalization to multiple targets. 

The radar tracks the target using a Bayesian tracker \cite{1998:YB}. The adversarial jammer
(mounted on a dedicated ECM ship \cite{2014:NR} or  aircraft) injects jamming power into the environment as an electronic countermeasure (ECM) to decrease the measurement accuracy of the radar. In response, the radar varies its pulse power as an electronic counter-countermeasure (ECCM) to mitigate the presence of the adversarial jammer.\footnote{In our setup, the jammer, and target are independent entities. \cite{2008:LX-et-al}, \cite{2018:CW-et-al}, \cite{2020:QL-et-al} also study a similar model with independent targets and jammers. In an alternative setup (not considered here), the jammer is mounted on the target \cite{2012:XS-et-al}. This results in the radar's observation $\radarObservation$ of the jamming power $\jammerStrategy$ also being dependent on the target's kinematics.}

It is convenient to formulate our setup in a three timescale framework as illustrated in Fig.~\ref{fig:block_diagram}. Let $\timeIndexOne\in \{0,1,2,\ldots\}$, $\timeIndexTwo\in \{0,1,2,\ldots\}$, and $\timeIndexThree\in\{0,1,2,\ldots\}$ denote the time index for fast, intermediate and slow timescale, respectively. On the fast timescale, the targets of interest evolve with linear, time-invariant dynamics and additive Gaussian noise. On the intermediate timescale, the radar and the jammer participate in an EW and update their ECCM and ECM, respectively. Finally, in the slow timescale, the target conducts maneuvers independent of the radar and the jammer. 

We consider the two frameworks below. Our first framework, discussed in 
 Sec.\ref{sec:fast_timescale} considers ECCM against barrage jamming (ECM) by an adversarial jammer; the jammer and the target operate independently. Our second framework discussed in Sec.\ref{sec:example-2} considers ECCM  against deception jamming by an adversarial jammer. Here the jammer exploits information about the target's state to jam the radar (for example, the jammer is mounted on the target). Finally, for both these frameworks, Sec.\ref{sec:intermediate_timescale} formulates the PAP for the optimal ECCM and ECM strategy of the radar and jammer.

\subsection{Example 1: ECCM against barrage jamming}
\label{sec:fast_timescale}
\subsubsection{Target dynamics}
\label{sec:target-dynamics}
The targets evolve on the fast time scale $\timeIndexOne$.  We use the standard linear Gaussian model\cite{1998:YB}  for the kinematics of the target and initial condition $x_0$:
\begin{align} 
	\label{eq:target_model}
	\begin{aligned}
		x_{0} &\sim \scrN(\hat{x}_0,\Variance_{x_0})\\ 
		x_{\timeIndexOne+1} &=A x_{\timeIndexOne}+w_{\timeIndexOne}\\
		A &=\operatorname{diag}\left[\left[\begin{array}{ll}1 & T \\ 0 & 1\end{array}\right],\left[\begin{array}{ll}1 & T \\ 0 & 1\end{array}\right],\left[\begin{array}{cc}1 & T \\ 0 & 1\end{array}\right]\right]
	\end{aligned}
\end{align}
Here, $T$ denotes the sampling time. The initial condition $\scrN(\hat{x}_0,\Variance_{x_0})$ denotes a Gaussian random vector with mean $\hat{x}_0$ and covariance $\Variance_{x_0}$. $x_\timeIndexOne\in\R^6$, comprised of the $x,y,z$ position and velocity, is the state of the target at time $\timeIndexOne$. The i.i.d. sequence of Gaussian random vectors $\{w_{\timeIndexOne}\sim\scrN(0,Q_t)\}$ models the acceleration maneuvers of the target. The covariance matrix $Q_t$ depends on the target maneuvers on the slow timescale.

\subsubsection{Radar's measurement model}
\label{sec:measurement-model}
The measurement vectors $y_{\timeIndexOne}$ of the target's state recorded at the radar are
\begin{align}
    \begin{aligned}
	\label{eq:measurement-eqn}
	&y_{\timeIndexOne} = h(x_k,v_k),\quad v_{\timeIndexOne}\sim\scrN\bigg(0,V\Big(\SNRn\Big)\bigg)
	\end{aligned}
\end{align}
Here, $h(\cdot)$ represents the radar's sensing functionalities. The measurement noise variance $V\Big(\SNRn\Big)$ of the random vector $v_k$ depends on  $\SNRn$ which will be defined in \eqref{eq:expected_SNR}. Here, $n$ indexes the intermediate timescale on which the radar and the jammer update their strategies\footnote{$\SNRn$ is the ratio of mean radiation pulse power of the radar to the mean jamming power observed by the radar, see \eqref{eq:expected_SNR}.}. The sequences $\{w_{\timeIndexOne}\}, \{v_{\timeIndexOne}\}$ are assumed to be statistically independent.

\subsubsection{Bayesian Tracker and Covariance}
We make the standard assumption that the radar has a Bayesian tracker \cite{1998:YB}, which recursively computes the posterior distribution $\Prob(x_k|y_1,\ldots y_k)$ of the target state at each time $k$.
For the nonlinear measurement  equation~\eqref{eq:measurement-eqn},
the posterior distribution can only be computed approximately, for example, using
a particle filter
\cite{2003:BR-SA-NG}.

\label{sec:bayesian-tracker}
In our PAP formulation  discussed in Sec.\ref{sec:intermediate_timescale},  we will require the covariance $\operatorname{Cov}(x_k|y_1,\ldots,y_k)$ of the posterior distribution of the target's state. 
The PAP is solved on the intermediate timescale. In comparison, the target evolves on the fast timescale. Therefore, in the PAP formulation below, we only need  the asymptotic value of the covariance, i.e., $\Variance := \lim_{k\rightarrow\infty}\operatorname{Cov}(x_k|y_1,\ldots,y_k)$. This can be estimated (in general) using a particle filter algorithm \cite{2003:BR-SA-NG}.

\subsubsection{Example. Linear measurement model and tracker}
\label{sec:linear-measurement-model}
A special case of  measurement equation~\eqref{eq:measurement-eqn}
is the linear case:
\begin{align}
	\label{eq:linear-measurement-equation}
	y_{\timeIndexOne} &=C x_{\timeIndexOne}+ \frac{v_{\timeIndexOne}}{\SNRn}
\end{align}
Then the posterior  $p(x_k|y_1,\ldots,y_k)$ is computed exactly by the Kalman filter \cite{1979:BA-JM}, \cite{2016:VK}. As mentioned above, in the PAP  we are interested only in the covariance of the posterior distribution. The Kalman filter equations provide a closed-form, recursive relation to compute the covariance: $\Variance_{k|k}:=\Expect{(x_k-\Expect{x_k|y_1,\ldots,y_k})^2}$; this can be computed independent of the observations \cite{2016:VK}.
Moreover, assuming $[A,C]$ is detectable and $[A,\sqrt{Q}]$ is stabilizable guarantees that the covariance $\Variance_{k|k}$ converges to the limiting solution $\Variance^{(n)}$ given by the \textit{algebraic Riccati equation} (ARE)~\eqref{eq:are}. We use $\Variance^{(n)}$ to denote the solution of the ARE at the end of the time index $\timeIndexTwo$ in the intermediate timescale. 
\begin{align}
	\label{eq:are}
	&\Variance^{(\timeIndexTwo)}-A\left(\Variance^{(\timeIndexTwo)}-\Variance^{(\timeIndexTwo)} C^{\prime}D_{\timeIndexTwo}^{-1} C \Variance^{(\timeIndexTwo)}\right) A^{\prime}-Q=0
\end{align}
where, $D_{\timeIndexTwo}:=\bracketSquare{C \Variance^{(\timeIndexTwo)} C^{\prime}+\frac{1}{\SNRn}}$.
The solution of ARE, i.e., $\Variance^{(\timeIndexTwo)}$, on the fast timescale parametrizes the PAP (described in Sec.\ref{sec:intermediate_timescale}). 

\subsubsection{Multiple targets}
\label{sec:multiple-targets} Thus far, we assumed that the jammer adversely affects the radar measurements of a single target. More generally, the jammer can affect the radar measurements of multiple targets.
If the jammer interferes with the radar estimates of multiple targets, the parameter $\maxEV{\Variance}$ appearing in the PAP (see Sec.\ref{sec:intermediate_timescale} below) can be modified by substituting a convex combination of the covariance of posterior distribution of the targets' state as in~\eqref{eq:weighted-covariance}, namely,
\begin{align}
	\label{eq:weighted-covariance}
	\begin{aligned}
		&\maxEV{\Variance} = \sum_{i=1}^{H} \beta_i\, \maxEV{\Variance_i}\\
		&\text{where, }\sum_{i=1}^{H}\beta_i = 1, \text{and } \beta_i \geq 0,\; \forall\, i\in\{1,2,\ldots,H\}\\
	\end{aligned}
\end{align}
Here $H$ denotes  the number of targets whose radar  measurements are adversely affected by the jammer. This can be implemented by replicating $H$ copies of the ARE~\eqref{eq:are}. Let us denote them as $\operatorname{ARE}_1, \operatorname{ARE}_2,\ldots, \operatorname{ARE}_H$ respectively. $\operatorname{ARE}_i$ solves the algebraic Riccati equation for the $i^{th}$ target and outputs $\Variance^{(n)}_i$ at the start of the the time index $\timeIndexTwo$ in the intermediate timescale. Substitute $\Variance^{(n)}_1, \Variance^{(n)}_2,\ldots,\Variance^{(n)}_H$ in~\eqref{eq:weighted-covariance} to obtain $\maxEV{\Variance}$.

\subsection{Example 2. ECCM against deception jamming}
\label{sec:example-2}
In the above section,  the jammer injected noise power into the environment to degrade the measurement accuracy of the radar; and the radar optimized its pulse power as an ECCM against the jammer's ECM. We assumed that the jammer's ECM was independent of the target's state. In this section, we consider a deception jamming framework. The jammer purposefully injects interference that depends on the state (position) of the target. (For example, the jammer is mounted on a target.) This designed interference introduced by the jammer affects the measurement accuracy of the radar as a function of the target's state.

We model deception jamming  as follows (recall $k,n,t$ index the fast, intermediate and the slow timescales):
\begin{align}
\label{eq:target_model_updated}
    \begin{aligned}
        x_{0} &\sim \scrN(\hat{x}_0,\Variance_{x_0}),\quad z_{0}\sim\scrN(\hat{x}_0,\Variance_{z_0})\\
        \begin{bmatrix}
            x_{k+1}\\
            z_{k+1}
        \end{bmatrix}&= \begin{bmatrix}
            A & 0\\
            B_1 & B_2
        \end{bmatrix}\begin{bmatrix}
            x_k\\
            z_k
        \end{bmatrix}+\begin{bmatrix}
            w_k\\
            v_k
        \end{bmatrix}\\
        y_k &= \begin{bmatrix}C_1 & C_2\end{bmatrix} \begin{bmatrix}
            x_k\\z_k
        \end{bmatrix}
    \end{aligned}
\end{align}

Here, matrix $A$ is as in \eqref{eq:target_model}. $z_k$ denotes the designed interference injected by the jammer at time $k$. The key point is that the designed interference $z_k$ depends on the target state $x_k$. For analytical tractability, we assume that $z_k$ is a linear function of state  $x_k$. Notice that $z_k$ affects the measurement of the radar through \eqref{eq:target_model_updated}, i.e., the jammer injects an additional component $Dz_k$  to the radar's measurement. 
The i.i.d. sequence of random vectors $\{w_k\sim \scrN(0,Q_t)\}$ models the acceleration maneuvers of the target. The covariance matrix $Q_t$ is updated by a target maneuver on the slow timescale. The i.i.d. random vectors $\{v_k\sim\scrN(0,\frac{1}{\SNRn})\}$ controls the impact of deception jamming on the measurement accuracy of the radar. $\SNRn$ is defined in \eqref{eq:expected_SNR}. It is a function of the radar's pulse power (ECCM) and the jamming power (ECM). 

In our PAP formulation  discussed in Sec.\ref{sec:intermediate_timescale} below,  we will require the covariance $\operatorname{Cov}(x_k|y_1,\ldots,y_k)$ of the posterior distribution of the target's state.  The PAP is solved on the intermediate timescale. In comparison, the target evolves on the fast timescale. Therefore, in the PAP formulation, we only need  the asymptotic value of the covariance, i.e., $\Variance=\Variance_{xx} := \lim_{k\rightarrow\infty}\operatorname{Cov}(x_k|y_1,\ldots,y_k)$. For the linear deception jamming model \eqref{eq:target_model_updated}, we  can compute $\Variance$ using the ARE \eqref{eq:are-modified} as
\begin{align}
    \label{eq:are-modified}
    \begin{aligned}
    &\Variance^{(\timeIndexTwo)}-\bar{A}\left(\Variance^{(\timeIndexTwo)}-\Variance^{(\timeIndexTwo)} \begin{bmatrix}C_1 \\ C_2\end{bmatrix} D_{\timeIndexTwo}^{-1} \begin{bmatrix}C_1 \\ C_2\end{bmatrix}^{\prime}  \Variance^{(\timeIndexTwo)}\right) \bar{A}^{\prime}-\bar{Q}=0\\
        &\bar{A}:=\begin{bmatrix}
            A & 0\\
            B_1 & B_2
        \end{bmatrix},\quad\bar{Q}:=\begin{bmatrix}
            Q & 0\\0 & \frac{1}{\SNRn}
        \end{bmatrix}\\
        &D_{\timeIndexTwo}:=\begin{bmatrix}C_1, C_2\end{bmatrix}^{\prime}\Variance^{(\timeIndexTwo)} \begin{bmatrix}C_1 \\ C_2\end{bmatrix},\quad\Variance^{(n)}=\begin{bmatrix}\Variance_{xx} & \Variance_{xz} \\ \Variance_{xz} & \Variance_{zz}\end{bmatrix}
    \end{aligned}
\end{align}
Here, $\Variance^{(n)}$ denotes the asymptotic covariance $\Variance$ at the intermediate time index $n$.

\subsubsection*{Nonlinear Model for  Deception Jamming}
The above linear model  is for analytical tractability. A more general deception jamming  model is $z_{k+1}=f(z_k,x_k,v_k)$ where  the designed interference $z_k$   impacts radar measurement as $y_k=h(x_k,z_k)$. Under suitable ergodicity conditions,  the asymptotic value of the covariance $\Variance=\Variance_{xx} := \lim_{k\rightarrow\infty}\operatorname{Cov}(x_k|y_1,\ldots,y_k)$ can be estimated using sequential Markov chain Monte-Carlo methods~\cite{2003:BR-SA-NG}. Recall that $k$ indexes the fast timescale. The PAP formulation of this paper (discussed below) also holds for such nonlinear  models in terms of $\Variance$.


\subsection{Principal-agent problem (PAP)}
\label{sec:intermediate_timescale}
Thus far, we described the dynamics of ECCM on three timescales. We now describe how the radar optimizes its ECCM strategy on the intermediate timescale as the solution of a PAP\footnote{There are two types of PAP \cite{1995:AM} studied in microeconomics: moral hazard and adversarial selection. In PAP with moral hazard, the principal receives a noisy signal of the agent's action and writes a contract to induce an agent's action that maximizes its utility.}. At the core of the PAP lies information asymmetry: the principal only views the agent's action in noise, so the principal needs to write a contract with suitable incentives to induce action from the agent that would maximize its utility. 

In our formulation, we treat the radar as the principal and the jammer as the agent. The EW setup between the radar and jammer is illustrated schematically in Fig.~\ref{fig:block_diagram}. The radar aims to track a target with kinematics described in~\eqref{eq:target_model}. The jammer performs ECM to decrease the SNR of the radar's measurement, and the radar observes a noisy signal~\eqref{eq:linear-measurement-equation} of the jammer's ECM. The radar then chooses its ECCM as a function of the noisy signal of the jammer's ECM to maximize its utility. 

We described our two frameworks for jamming, namely barrage jamming and deception jamming in Sec.\ref{sec:fast_timescale} and Sec.\ref{sec:example-2}, respectively. In both frameworks, ECM and ECCM impact the measurement accuracy of the radar through the random process $\{v_k\}$. The jammer controls its jamming power (ECM) to increase the variance of $\{v_k\}$ whereas the radar controls its radiation pulse power (ECCM) to decrease the variance of $\{v_k\}$. Hence, the PAP is the same for barrage and deception jamming.


\subsubsection{ECCM as a PAP}
\label{sec:eccm-pap}
We assume the jamming power $\jammerStrategy$ of the jammer takes values from the finite set $\jammerStrategyOptionSet$:
\begin{equation}
	\label{eq:jammer-strategy-set}
\jammerStrategyOptionSet =  		\{\jammerStrategyOption_1,\jammerStrategyOption_2,\ldots,\jammerStrategyOption_M\},\;\jammerStrategyOption_1\leq\jammerStrategyOption_2\leq\ldots\leq\jammerStrategyOption_M
\end{equation} 
Below we denote random variables in boldface.
Due to the measurement noise~\eqref{eq:linear-measurement-equation}, radar observes a noisy signal $\boldRadarObservation\in\jammerStrategyOptionSet$ when the jamming power is $\jammerStrategy$. The conditional pmf $\condPMF{\radarObservation}{\jammerStrategy}$ gives the probability that radar observes $\radarObservation$ given the jamming power is $\jammerStrategy$.
We have modeled the observation uncertainty $\condPMF{\radarObservation}{\jammerStrategy}$ using a time-invariant, memoryless channel \cite{2006:TC-TJ}. This inherently assumes that the ambient condition, including obstacles, source of electromagnetic noise, are constant throughout the EW. One can use a more sophisticated model to include moving obstacles and time-varying sources of electromagnetic noise. We deal with a time-invariant, memoryless model for simplicity. 

The radar observes the jamming power $\boldRadarObservation$ and chooses a radiation pulse power (ECCM), $\radarStrategyRandom$, to maximize its utility in the presence of the jammer's ECM.
Let $\radarStrategy\in\R^M$  denote the radar's ECCM strategy for each possible observation of the jammer: 
\begin{align}
    \begin{aligned}
        \label{eq:e_R}
        &\radarStrategy_i=\operatorname{e}_{\jammerStrategyOption_i}^\prime \radarStrategy,\quad 
	    &\operatorname{e}_{\jammerStrategyOption_i}:=e_i,\;\forall i\in\{1,2,\ldots,M\}
    \end{aligned}
\end{align}
Here, $e_i$ denotes the standard unit vector in $\R^M$ with $i^{th}$ entry as 1; and $e_i^\prime$ denotes its transpose. 

{\em Remark.}
We assume the jammer knows the radar's channel model parameters $\condPMF{\radarObservation}{\jammerStrategy}$. 
For  the linear measurement model~\eqref{eq:linear-measurement-equation}, the assumption implies that  the radar and jammer compute  the same target covariance $\Variance$ via ARE~\eqref{eq:are}. 
In Sec.\ref{sec:numerical-result-noisy-channel-estimate}, we give a detailed discussion and numerical examples when the jammer has imperfect information about the radar's channel. We discuss how this imperfect information affects the optimal ECCM strategy.

We can now formulate the radar's ECCM problem as a PAP. The PAP for both barrage and deception jamming is the following constrained optimization problem:
\begin{subequations}
\label{eq:pa-general}
\begin{empheq}[box=\widefbox]{align}
	\label{eq:pa-general-objective}
	&\max_{\substack{\jammerStrategy\in\jammerStrategyOptionSet,\\\pi\in\R^M}}\; \radarUtility(\radarStrategy,\jammerStrategy)\\
	\label{eq:pa-general-ic}
	&\text{s.t.}\;\arg\max_{\bar{J}\in\jammerStrategyOptionSet}\,\jammerUtility(\radarStrategy,\bar{\jammerStrategy})=\jammerStrategy
	\end{empheq}
\end{subequations}
Here, $\radarUtility(\radarStrategy,\jammerStrategy)$, $\jammerUtility(\radarStrategy,\jammerStrategy)$  denote the utility functions of the radar and the jammer respectively:
\begin{subequations}
    \label{eq:utility-functions}
    \begin{align}
    	\label{eq:radar-utility-function}
    	&\radarUtility(\radarStrategy,\jammerStrategy):=\mathbb{E}_{\radarObservation|\jammerStrategy}\left[\maxEV{\Variance\indn}\,c_1\,\log\bracketRound{\operatorname{SNR}(\boldRadarObservation)}-(\radarStrategyRandom)^2\right]
    	\\
    	\label{eq:jammer-utility-function}
    	&\jammerUtility(\radarStrategy,\jammerStrategy):=\mathbb{E}_{\radarObservation|\jammerStrategy}\left[\frac{1}{\maxEV{\Variance\indn}}c_2\log\bracketRound{\frac{1}{\operatorname{SNR}(\boldRadarObservation)}}\right]-\jammerStrategy^2
    \end{align}
\end{subequations}
$c_1,c_2$ are positive constants, and  $\operatorname{e}_{\boldRadarObservation}$ is defined in~\eqref{eq:e_R}. $\operatorname{SNR}(\boldRadarObservation)$ in \eqUTILITYfunctions\ denotes the measurement SNR of the radar:
\begin{align}
	\label{eq:SNR}
	\begin{aligned}
		\operatorname{SNR}(\boldRadarObservation)&:=\frac{\radarStrategyRandom}{\boldRadarObservation}
	\end{aligned}
\end{align}
The first term, $c_1\log\bracketRound{\operatorname{SNR}(\boldRadarObservation)}$ in~\eqref{eq:radar-utility-function} is the radar's reward to improve its measurement's SNR. The logarithm of SNR as a candidate function is closely related to channel capacity for an analog channel subject to additive white Gaussian noise \cite{2006:TC-TJ}. Hence, we substitute it for our reward functions given its practical significance. The first term, $c_2\log\bracketRound{\frac{1}{\operatorname{SNR}(\boldRadarObservation)}}$, in~\eqref{eq:jammer-utility-function} is the jammer's reward to decrease the measurement's SNR~\eqref{eq:SNR} of the radar. It captures the adversarial nature  of the jammer: while the radar's reward is  increasing in  $\log(\operatorname{SNR}(\boldRadarObservation))$, the jammer aims  to minimize $\log(\operatorname{SNR}(\boldRadarObservation))$. The second term, $(\radarStrategyRandom)^2$, in~\eqref{eq:radar-utility-function} is the radar's cost for using a pulse power $\radarStrategyRandom$ in response to the observed jamming power $\boldRadarObservation$. The second term, $\jammerStrategy^2$, in~\eqref{eq:jammer-utility-function} is the cost for injecting the jamming power $\jammerStrategy$. Choosing a convex, increasing function for cost in~\eqref{eq:radar-utility-function}-\eqref{eq:jammer-utility-function} models an increasing cost function whose marginal value increases at higher effort.

$\maxEV{\Variance}$ in \eqUTILITYfunctions\  denotes the maximum eigenvalue of the covariance matrix $\Variance$.
It is a useful scalar-valued measure\footnote{Another useful measure is the trace of the covariance.  Our framework allows for any scalar-valued measure $\maxEV{\Variance}$ for the covariance of the posterior distribution of the target's state.} of the covariance of the posterior distribution of the target's state. $\maxEV{\Variance}$ parametrizes the reward function for the radar and the jammer. To be specific, it scales the reward component of the radar's and the jammer's utility function \eqUTILITYfunctions. This implies: when the uncertainty in the target's maneuver is large, the radar's reward $\maxEV{\Variance}c_1 \log(\operatorname{SNR}(\boldRadarObservation))$ for a higher SNR increases with an increase in target's maneuver. It incentivizes the radar to aim for better measurement accuracy when the target's maneuver is large. For the jammer, it's the other way round: the jammer's reward $\frac{1}{\maxEV{\Variance\indn}}c_2\log\bracketRound{\frac{1}{\operatorname{SNR}(\boldRadarObservation)}}$ for a higher SNR decreases with increase in target's maneuver. It ensures that the jammer does not waste its effort to decrease the measurements' accuracy of the radar when the target maneuver is large.

At first sight, the PAP~\eqrefPAgeneral\ is a mixed-integer program. The objective~\eqref{eq:pa-general-objective} is the radar's utility function. The incentive constraints~\eqref{eq:pa-general-ic} incentivizes the jammer to take the action $\jammerStrategy$. To obtain the optimal ECCM strategy, the radar first solves the PAP~\eqrefPAgeneral\ for each jamming power $\bar{\jammerStrategy}\in\jammerStrategyOptionSet$. The radar then incentivizes the jamming power $\jammerStrategy$, which yields the best utility. To efficiently  compute the radar's ECCM strategy for a fixed $\jammerStrategy$, we will derive an equivalent convex optimization problem for the PAP~\eqrefPAgeneral\ in Sec.\ref{sec:convex_optimization} below. This facilitates the use of convex optimization solvers for the radar's ECCM problem.

\subsubsection{Discussion of utility functions for the radar and the jammer}
We discussed the PAP~\eqrefPAgeneral\ for a specific choice of the utility functions defined in \eqUTILITYfunctions. The utility function was linearly separable into a reward for improving the performance and a cost for choosing a particular strategy. One can consider more general utility functions for the radar and  jammer:
\begin{align}
	\label{eq:utility-function-general}
	\begin{aligned}
		\radarUtility(\radarStrategy,\jammerStrategy)&:=\mathbb{E}_{\radarObservation|\jammerStrategy}\left[\maxEV{\Variance\indn}\,g_1\bracketRound{\operatorname{SNR}(\boldRadarObservation)}-g_2(\radarStrategyRandom)\right]\\
		\jammerUtility(\radarStrategy,\jammerStrategy)&:=\mathbb{E}_{\radarObservation|\jammerStrategy}\left[\frac{1}{\maxEV{\Variance\indn}}f_1\bracketRound{\operatorname{SNR}(\boldRadarObservation)}\right]-f_2(\jammerStrategy)
	\end{aligned}
\end{align}
Here, $\operatorname{e}_{\radarObservation}$ is defined in~\eqref{eq:e_R}. $\operatorname{SNR}(\boldRadarObservation)$ is defined in~\eqref{eq:SNR}. $g_1(\cdot),f_1(\cdot)$ are increasing, concave functions that model  the reward of the radar and the jammer respectively; they model the radar and the jammer as a risk-averse agents. The functions $g_2(\cdot),f_2(\cdot)$ are increasing, convex functions and model  the cost of the radar and  jammer, respectively. The convexity assumption  ensures that the marginal cost increases at higher effort.

\subsubsection{Convex optimization formulation for  radar's ECCM problem}\label{sec:convex_optimization}
In  PAP~\eqrefPAgeneral, the radar optimizes its ECCM strategy $\radarStrategy$ for each jamming power $\bar{\jammerStrategy}\in\jammerStrategyOptionSet$ and then incentivizes a jamming power $\jammerStrategy$ that yields it maximum utility. Our main result below is to  construct an equivalent convex optimization problem for the PAP~\eqrefPAgeneral\ for any fixed jamming power $\jammerStrategy$:
\begin{theorem}
	\label{thm:convex}
	For any fixed jamming power $\jammerStrategy$, the PAP~\eqrefPAgeneral\ for the radar's ECCM problem is equivalent to the following convex optimization problem in $x$ (recall $e_\boldRadarObservation$ is defined in~\eqref{eq:e_R}):
	\begin{subequations}
	\label{eq:pa-sc-simplified}
	\begin{align}
		\label{eq:pa-sc-simplified-objective}
        &\max_{\substack{x\in\R^M}}\; \mathbb{E}_{\radarObservation|\jammerStrategy}[c_1\,\maxEV{\Variance} \,(\radarStrategyRandomX - \log(\boldRadarObservation)) - \exp(2\,\radarStrategyRandomX)]\\
		\intertext{subject to the following affine constraint on $x$ for a fixed jamming power $J$:}
		\label{eq:pa-sc-simplified-ic}
		&\arg\max_{\bar{\jammerStrategy}\in\jammerStrategyOptionSet} \mathbb{E}_{\radarObservation|\bar{\jammerStrategy}}\bracketSquare{\frac{c_2}{\maxEV{\Variance}}\,\bracketRound{\log(\boldRadarObservation)- \radarStrategyRandomX}}-\bar{\jammerStrategy}^{2}=\jammerStrategy
	\end{align}	
	\end{subequations}
    Finally, given the solution $x^*$ to  PAP~\eqrefPAscSIMPLIFIED, the  radar's optimal ECCM strategy $\radarStrategy^*$  is 
	\begin{align}
		\label{eq:radar-strategy-x-relation}
		\radarStrategy^*_i = \exp(x^*_i),\, i\in\{1,2,\ldots,M\}
	\end{align}
\end{theorem}
{\em Remark}.
The intuition behind  Theorem~\ref{thm:convex} is as follows. The objective~\eqref{eq:pa-sc-simplified-objective} is a sum of affine and a concave function in $x$; the incentive constraints~\eqref{eq:pa-sc-simplified-ic} can be equivalently represented as a set of affine inequality constraints.

Let us denote the solution of the PAP~\eqrefPAscSIMPLIFIED\ at the time instant $n$ as ($\radarStrategy^{(n)},\jammerStrategy^{(n)})\in\R^M\times\jammerStrategyOptionSet$. The solution of the ARE~\eqref{eq:are} at time index $n$ can be computed using the expected signal-to-noise ratio $\overline{\operatorname{SNR}}^{(n)}$:
\begin{align}
	\label{eq:expected_SNR}
	\SNRn=\frac{\mathbb{E}_{\radarObservation|\jammerStrategy}\left[\operatorname{e}_\boldRadarObservation^\prime\radarStrategy^{(n)}\right]}{\mathbb{E}_{\radarObservation|\jammerStrategy}\left[\boldRadarObservation\right]}
\end{align}

\subsection*{Radar-jammer interaction as a non-cooperative game} We have formulated the radar's ECCM problem as a  convex optimization problem (Theorem~\ref{thm:convex}). The formulation also yields useful structural results, as will be discussed in Sec.\ref{sec:structural-results}. An alternative approach is to formulate a non-cooperative dynamic game and characterize the perfect Bayesian Nash equilibrium. For the utility functions specified in \eqref{eq:utility-functions}, it turns out that we have a  supermodular game \cite{2011:DT}\cite{2005:RA} with the radar and jammer as players. For a supermodular game, it is well known that a pure strategy Nash equilibrium exists, but there could be multiple Nash equilibria.  If the supermodular game has a unique Nash equilibrium, it can be computed using best-response dynamics \cite{2011:DT}\cite{2005:RA}. In general, additional constraints have to be imposed on the utility functions to guarantee uniqueness. One can also formulate a Stackelberg game to obtain the radar's optimal ECCM strategy. With the radar as the leader and the jammer as the follower, the solution of PAP \eqrefPAscSIMPLIFIED\ is also a Nash equilibrium for the Stackelberg game. 
The advantage of the PAP framework is that it provides a convenient way to impose additional constraints on the problem. For example, suppose the radar wants to ensure that the utility of the jammer is below a certain threshold. This can be straightforwardly handled by adding an inequality constraint to the PAP \eqrefPAscSIMPLIFIED.

\subsection{ECM and ECCM Implementation details}
The target evolves kinematically on the fast timescale according to the dynamics in~\eqref{eq:target_model}. Its maneuver $Q_t$ is updated in the slow timescale, independent of the strategies of the radar and the jammer. Given the ECCM and ECM strategies $\radarStrategy^{(n)},\jammerStrategy^{(n)}$ at the time index $\timeIndexTwo$, the radar and the jammer solve the ARE~\eqref{eq:are} to obtain covariance $\Variance^{(n)}$ of the posterior distribution of the target's state. We can use the ARE \eqref{eq:are} to obtain the covariance $\Variance^{(n)}$ because the target evolves on the fast timescale and the radar's ECCM problem is solved at the intermediate timescale. $\Variance^{(n)}$ updates the utility functions~\eqref{eq:utility-functions} of the radar and the jammer. The radar then solves the radar's ECCM problem~\eqrefPAscSIMPLIFIED\ at time $n+1$ to find the optimal ECCM and ECM strategy $\radarStrategy^{(n+1)},\jammerStrategy^{(n+1)}$, respectively. This step requires solving $|\jammerStrategyOptionSet|$ convex optimization problems \eqrefPAscSIMPLIFIED. The optimal ECCM and ECM strategies converge before the next update of the target maneuver $Q_t$.


To summarize, this section formulated the radar's ECCM problem as a PAP. For specific choices of utility functions for the radar and the jammer, we formulated a convex optimization problem~\eqrefPAscSIMPLIFIED\ for the radar's ECCM problem for each $\jammerStrategy\in \scrJ$.
\section{Structure of the ECCM Strategy}
	\label{sec:structural-results}
	
	Although the previous section formulated ECCM as a convex optimization problem, a natural question is: {\em Does the problem have sufficient structure so that the optimal ECCM strategy can be characterized without brute force computation?} This section gives sufficient conditions for the radar's optimal ECCM strategy~\eqrefPAscSIMPLIFIED\ to be an increasing function of the observed jamming power. Our result is motivated by the structural result for  PAP with moral hazard \cite{1988:IJ}.
	
	The structural result in this section is useful for two reasons. First, it yields a useful heuristic for the choice of the optimal ECCM strategy; namely, it is an increasing function of the observed jamming power. Second,  if we constrain the solution of the PAP~\eqrefPAscSIMPLIFIED\ within the class of increasing affine functions\footnote{Since the optimal strategy is monotone, the monotone affine ECCM strategy obtained below qualifies as the optimal affine approximation to PAP.} (defined below  in~\eqref{eq:parametrize_soln}), then we can reduce search-space for the radar's ECCM strategy in~\eqrefPAscSIMPLIFIED\ from $\R^M$ to $\R^2$. This enables efficient computation of a constrained solution at a low computation cost. For example, consider computing the optimal solution for the constrained optimization problem~\eqrefPAscSIMPLIFIED\ using the projected gradient descent algorithm \cite{2004:SB-LV}. It requires the computation of gradient at each update step. The cost for calculating gradient is of the order $\scrO(M)$, where $M$ is the dimension of the search space. Hence, reducing the search-space from $\R^M$ to $\R^2$ can significantly reduce the computation for large  $M$.

To present our result, we first define a stochastic ordering \cite{2002:AM-DS} for the conditional pmf $\condPMF{\radarObservation}{\jammerStrategy}$. We represent the conditional pmf $\condPMF{\radarObservation}{\jammerStrategy}$ as a stochastic matrix $\probMatrix$ with elements
	\begin{align}
	    \label{eq:stochastic-matrix}
	    \probMatrix_{m \desired}:= \condPMF{\jammerStrategyOption_\desired}{\jammerStrategyOption_m}
	\end{align}
	\begin{definition}[Total positivity of order 2 (TP2) \cite{1980:SK-YR}]
		The stochastic matrix $\probMatrix$~\eqref{eq:stochastic-matrix} satisfies TP2 if $\forall\,i,j,m,n\in\{1,2,\ldots,M\}$ with $i>j, m>n$,
		\begin{align}
		\label{eq:tp2}
		\probMatrix_{i m} \probMatrix_{j n} \geq \probMatrix_{i n} \probMatrix_{j m}
		\end{align}
	\end{definition}
Now consider  a \textit{relaxed radar's ECCM problem} by modifying the constraint~\eqref{eq:pa-sc-simplified-ic}:
	\begin{definition}[Relaxed Radar's ECCM Problem]
		A relaxed radar's ECCM problem is obtained from~\eqrefPAscSIMPLIFIED\ by modifying the constraint~\eqref{eq:pa-sc-simplified-ic} on $x$ to ($\operatorname{e}_\boldRadarObservation$ below is defined in~\eqref{eq:e_R}):
		\begin{align}
		    \intertext{For a fixed jammer power $\jammerStrategy$, $x$ satisfies the affine constraint}
		    \label{eq:relaxed-pap}
		    \arg\max_{\bar{\jammerStrategy}\geq\jammerStrategy}\left(\mathbb{E}_{\radarObservation|\bar{\jammerStrategy}}\left[\frac{c_2}{\maxEV{\Variance}}\,(\log(\boldRadarObservation)- \radarStrategyRandomX)\right]-\bar{\jammerStrategy}^{2}\right)=\jammerStrategy
		\end{align}
	\end{definition}

 The set of $x$ satisfying the constraint~\eqref{eq:pa-sc-simplified-ic} is a subset of the set of $x$ satisfying the constraint~\eqref{eq:relaxed-pap}. Hence, we call optimization of~\eqref{eq:pa-sc-simplified-objective} subject to~\eqref{eq:relaxed-pap} as the relaxed radar's ECCM problem. 
	This implies that the maximum value of the relaxed radar's ECCM problem~\eqref{eq:relaxed-pap} is an upper bound to the radar's ECCM problem~\eqrefPAscSIMPLIFIED. Our main result in this section would establish that the maximum value of~\eqref{eq:relaxed-pap} is the same as that of the PAP~\eqrefPAscSIMPLIFIED\ under certain conditions. We have already described the TP2~\eqref{eq:tp2} requirement on the stochastic matrix $P$~\eqref{eq:stochastic-matrix}. We now describe the second requirement on the convexity of the tail distribution of the conditional pmf $\condPMF{\radarObservation}{\jammerStrategy}$:
	\begin{align}
	    \label{eq:tail-sum}
	    \Pi_{i}\bracketRound{\jammerStrategy}:=\displaystyle{\Prob\bracketRound{\boldRadarObservation \geq \jammerStrategyOption_{i} \mid \textbf{\jammerStrategy}=\jammerStrategy}},\;\forall i\in\{1,2,\ldots,M\}
	\end{align}
	
	We use convexity of $\Pi_{i}\bracketRound{\jammerStrategy},\;\forall i\in\{1,2,\ldots,M\}$~\eqref{eq:tail-sum} to prove the concavity of the jammer's utility function $\jammerUtility(\radarStrategy,\jammerStrategy)$~\eqref{eq:jammer-utility-function}:
	
	\begin{lemma}
		\label{lemma:jammer-utility-concave}
		Let $x^*$ be the solution of the relaxed radar's ECCM problem~\eqref{eq:relaxed-pap}. If $x^*_1\leq x^*_2 \leq \ldots \leq x^*_M$ and $\Pi_{i}\bracketRound{\jammerStrategy}$~\eqref{eq:tail-sum} is convex in $\jammerStrategy,\;\forall i\in\{1,2,\ldots\,M\}$, then for the relaxed radar's ECCM problem~\eqref{eq:relaxed-pap}, the jammer's utility $\jammerUtility(\radarStrategy^*,\jammerStrategy)$~\eqref{eq:jammer-utility-function} is concave in $\jammerStrategy$. 
	\end{lemma}
	Here, the relation between $\radarStrategy^*$ and $x^*$ is given by~\eqref{eq:radar-strategy-x-relation}. 
	We are ready to discuss our structural result concerning the monotonicity of the solution of the PAP~\eqrefPAscSIMPLIFIED\ for the radar's ECCM problem. The result provides sufficient conditions under which the radar's optimal ECCM strategy for the radar's ECCM problem~\eqrefPAscSIMPLIFIED\ is an increasing function of the radar's observation of the jamming power.
	\begin{theorem}
		\label{thm:monotone}
		If the stochastic matrix $\probMatrix$~\eqref{eq:stochastic-matrix} is TP2~\eqref{eq:tp2} then for any choice of $\jammerStrategy$, solution of the relaxed radar's ECCM problem~\eqref{eq:relaxed-pap} is non-decreasing, i.e., $x_1^*\leq x_2^* \leq \ldots \leq x_M^*$. Moreover, if $\Pi_{i}\bracketRound{\jammerStrategy}$ is convex in $\jammerStrategy,\;\forall i\in\{1,2,\ldots,M\}$ then any solution of the relaxed radar's ECCM problem~\eqref{eq:relaxed-pap} is a solution to the radar's ECCM problem~\eqrefPAscSIMPLIFIED, i.e., there is an radar's optimal ECCM strategy to the radar's ECCM problem~\eqrefPAscSIMPLIFIED\ s.t. $\radarStrategy_1^*\leq \radarStrategy_2^* \leq \ldots \leq \radarStrategy_M^*$.
	\end{theorem}
	Here, the relation between $\radarStrategy^*$ and $x^*$ is given by~\eqref{eq:radar-strategy-x-relation}. In the context of the radar's ECCM problem, TP2~\eqref{eq:tp2} requirement on the matrix $\probMatrix$~\eqref{eq:stochastic-matrix} implies that 
	if the jammer uses a high jamming power $\jammerStrategy$, the radar's noisy observation of the jammer's action $\radarObservation$ is likely higher. This is a reasonable assumption for additive ambient noise (discussed in Sec.\ref{sec:eccm-pap}). For a complex radar's channel model, one can find the closest approximation of $\condPMF{\radarObservation}{\jammerStrategy}$ within the class of TP2 conditional distribution using a suitable statistical distance, e.g., total variation distance \cite{2003:CV}. This yields a sub-optimal ECCM strategy for the radar with a reduced computation cost.
	The convexity of the tail distribution $\Pi_i(\jammerStrategy),\;\forall i\in\{1,2,\ldots,M\}$~\eqref{eq:tail-sum} is used in Lemma~\ref{lemma:jammer-utility-concave} to conclude that the jammer's utility $\jammerUtility(\radarStrategy^*,\jammerStrategy)$~\eqref{eq:jammer-utility-function} is concave in $\jammerStrategy$. Concave utility function implies diminishing marginal return with increasing effort. This is frequent in real life. We make a weaker assumption about the convexity of the tail distribution $\Pi_{i}\bracketRound{\jammerStrategy},\;\forall i\in\{1,2,\ldots,M\}$~\eqref{eq:tail-sum} to conclude the concavity of the jammer's utility function $\jammerUtility(\radarStrategy^*,\jammerStrategy)$~\eqref{eq:jammer-utility-function}. Hence, it is also a well-justified assumption.
	
	Theorem~\ref{thm:monotone} facilitates a constrained solution to the radar's ECCM problem~\eqrefPAscSIMPLIFIED\ by restricting the search space for $x$ within the class of increasing affine functions:
	\begin{align}
    	\label{eq:parametrize_soln}
    	x_i = c_3 \jammerStrategyOption_i + c_4;\;c_3\in\R_{\small +}, c_4\in\R,\;i\in\{1,2,\ldots,M\}
	\end{align}
	The parametrization~\eqref{eq:parametrize_soln} reduces the search-space for $x$ from $\R^M$ to $\R^2$. In Sec.\ref{sec:rl-eccm}, we apply Theorem~\ref{thm:monotone} to compute a constrained solution to the radar's ECCM problem~\eqrefPAscSIMPLIFIED.
\section{Numerical Examples of PAP based  ECCM}
This section illustrates via numerical examples the application of the PAP~\eqrefPAscSIMPLIFIED\ for the radar's ECCM problem. Sec.\ref{sec:model_setup} summarizes the ECCM model setup involving a radar, a target and a jammer. It also specifies the model parameter and simulates the model. In Sec.\ref{sec:rl-eccm}, we exploit Theorem~\ref{thm:monotone} to parametrize the solution of the radar's ECCM problem; it enables computation of the radar's optimal ECCM within the class of increasing affine functions at a low computation cost.
	\label{sec:simulation}
	
	\subsection{Radar's optimal ECCM strategy for barrage jamming (Sec.\ref{sec:fast_timescale}) using the PAP}
	\label{sec:model_setup}
	Recall the schematic setup  in Fig.~\ref{fig:block_diagram}. A target evolves kinematically on the fast timescale according to the dynamics in~\eqref{eq:target_model}; the target is maneuvered on the slow timescale. The radar's measurement equation is a linear function with additive Gaussian noise~\eqref{eq:linear-measurement-equation}. The radar tracks the target, and a jammer attempts to decrease the radar's measurement accuracy by injecting jamming power as an ECM. The radar only observes the jamming power in noise. In order to operate efficiently, the radar is forced to vary its pulse power to probe the target (ECCM) as a function of the observed jamming power to mitigate the impact of ECM. The radar's optimal ECCM is solved using the PAP~\eqrefPAscSIMPLIFIED.

\begin{figure}[ht]
	    \begin{center}
	    \scalebox{0.8}{
	    \begin{tikzpicture}
	        \tikzstyle{process} = [rectangle, minimum height=1cm, text centered, draw=black];
	        \tikzstyle{arrow} = [thick,->,>=stealth];
	        \node(are0)[process]{ARE};
	        \node(pap1)[process, right of=are0, xshift=25]{PAP};
	        \node(are1)[process, right of=pap1, xshift=25]{ARE};
	        \node(pap2)[process, right of=are1, xshift=25]{PAP};
	        \node(inf)[process, right of=pap2, xshift=25, draw=none]{...};
	        \draw [arrow] (are0) -- node[anchor=south] {$\Variance^{(0)}$} (pap1);
	        \draw [arrow] (pap1) -- node[anchor=south] {$\radarStrategy^{(1)}$} node[anchor=north] {$\jammerStrategy^{(1)}$} (are1);
	        \draw [arrow] (are1) -- node[anchor=south] {$\Variance^{(1)}$} (pap2);
	        \draw [arrow] (pap2) -- node[anchor=south] {$\radarStrategy^{(2)}$} node[anchor=north] {$\jammerStrategy^{(2)}$} (inf);
	    \end{tikzpicture}
	    }
	    \end{center}
	    \caption{Schematic interaction between the radar and the jammer for a fixed target maneuver. ARE denotes the algebraic Riccati equation~\eqref{eq:are}; PAP refers to the PAP~\eqrefPAscSIMPLIFIED\ for the radar's ECCM problem. The radar computes the asymptotic covariance $\Variance^{(n)}$ of the target's posterior distribution using the ARE~\eqref{eq:are}. The covariance $\Variance^{(n)}$ paramterizes the radar's ECCM problem~\eqrefPAscSIMPLIFIED. The radar's ECCM problem~\eqrefPAscSIMPLIFIED\ is used by the radar to solve the radar's optimal ECCM strategy $\radarStrategy^{(n+1)}$; it also incentivizes the jammer to choose the jamming power $\jammerStrategy^{(n+1)}$. The solution pair $(\radarStrategy^{(n+1)},\jammerStrategy^{(n+1)})$ is then used to compute the $\Variance^{(n+1)}$ using the ARE~\eqref{eq:are}. We repeat the computations till the equilibrium is reached.}
	    \label{fig:ew-progress}
	\end{figure}
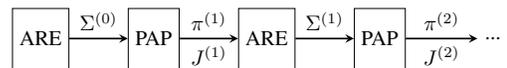

	We now simulate the PAP~\eqrefPAscSIMPLIFIED\ for finding the radar's optimal ECCM strategy for the target tracking problem detailed above. The model parameters are tabulated in the Table \ref{tab:simulation_parameters}. 
	\def\arraystretch{1.5}
	\begin{table}
		\centering
		\caption{Simulation Parameters for the Radar's ECCM Problem against Barrage Jamming.}
		\begin{tabular}{r|r|l}
			\hline
			\textbf{Parameters} & \textbf{Eq.} & \textbf{Value}\\
			\hline
			$T$ &~\eqref{eq:target_model}& 1\\
			$Q_0$ &~\eqref{eq:target_model}& $I_{6\times 6}$\\
			$C$ &~\eqref{eq:linear-measurement-equation} & $I_{6\times 6}$\\
			$\{\jammerStrategyOption_1,\ldots,\jammerStrategyOption_4\}$ &~\eqref{eq:jammer-strategy-set} & $\{1,2,3,4\}$\\
			$c_1$ &~\eqref{eq:pa-sc-simplified-objective} & $1\times10^{2}$\\
			$c_2$ &~\eqref{eq:pa-sc-simplified-ic} & $1\times10^{4}$\\
			$\probMatrix$ &~\eqref{eq:stochastic-matrix} & $\begin{bmatrix}
			0.3878 & 0.3215 & 0.1858 & 0.1049\\
			0.2980 & 0.3617 & 0.2146 & 0.1256\\
			0.2040 & 0.2583 & 0.3307 & 0.2070\\
			0.1029 & 0.1408 & 0.2140 & 0.5422
			\end{bmatrix}$\\
			\hline
		\end{tabular}
		\label{tab:simulation_parameters}
	\end{table}
	We simulated our model for a horizon length of $4$ on the slow timescale with maneuvers updated as $Q_{\timeIndexThree}=\timeIndexThree Q_0$ in the slow timescale. One unit of time in the slow timescale is chosen to be 8 units in the intermediate timescale. This allows sufficient time for the transients in the intermediate timescale to settle down. A block diagram of the interaction between the radar and jammer for a fixed target maneuver is shown in Fig.~\ref{fig:ew-progress}.
	
\begin{figure}
		\centering
		\includegraphics[scale=0.4]{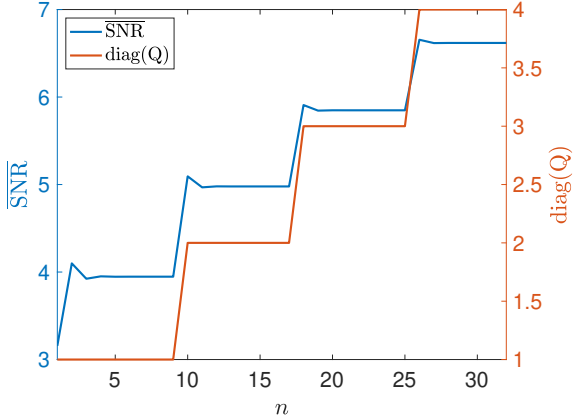}
		\caption{Expected signal-to-noise ratio ($\overline{\operatorname{SNR}}$)~\eqref{eq:expected_SNR} vs. intermediate time index ($n$) for barrage jamming. The parameters are specified  in Table \ref{tab:simulation_parameters}. Additionally, the target maneuver is updated as $Q_t=\timeIndexThree Q_0$ in the slow timescale. For each maneuver in the slow timescale, simulation was run for a horizon length=8 in the intermediate timescale. As state noise covariance $Q$ increases, the radar's marginal reward increases, whereas the marginal cost to increase the effort remains the same. For the jammer, an increase in state noise covariance $Q$ decreases the marginal reward, whereas the marginal cost to increase the effort remains the same. Hence, overall the SNR of the radar improves with an increase in state noise covariance $Q$.}
		\label{fig:SNR}
	\end{figure}

	The simulation is initialized with $\Variance^{(0)}=\maxEV{Q_0}=1$ as the solution of ARE~\eqref{eq:are}.
	The output $\overline{\operatorname{SNR}}$ vs time $n$ is displayed in Fig.~\ref{fig:SNR}. $Q_t=t I_{6\times 6}$ and $\operatorname{diag}(Q)$ represents the diagonal element of $Q$.
	
  A well-known property of the ARE of the Kalman filter is that as the state noise covariance $Q$ increases, so does $\maxEV{\Variance}$ \cite{1979:BA-JM}. Therefore, as state noise covariance $Q$ increases, the radar's marginal reward increases, whereas the marginal cost to increase the effort remains the same. For the jammer, an increase in state noise covariance $Q$ decreases the marginal reward, whereas the marginal cost to increase the effort remains the same. Hence,  the SNR of the radar increases with an increase in the target's state noise covariance $Q$ as shown in Fig.~\ref{fig:SNR}.
	
	\subsection{Constrained ECCM strategies for barrage jamming (Sec.\ref{sec:fast_timescale}) that exploit the PAP structure}
	\label{sec:rl-eccm}
	In Sec.\ref{sec:structural-results} we showed that the radar's ECCM strategy is an increasing function of the observed jamming power. We now exploit this property to construct a constrained ECCM strategy by the radar that is computationally efficient.
	To motivate this,  if the cardinality of the jamming noise power, $|\scrJ|$, is large, then computing the optimal solution to the PAP~\eqrefPAscSIMPLIFIED\ is not tractable for real-time computation. This is because the radar has to solve $|\scrJ|$  convex optimization problems. We now exploit Theorem~\ref{thm:monotone} to approximate the radar's optimal ECCM strategy.
	The model parameters specified in Table \ref{tab:simulation_parameters} satisfies the conditions in Theorem~\ref{thm:monotone}. Therefore, we parametrize the radar's ECCM strategy as an increasing affine function~\eqref{eq:parametrize_soln} of the observed jamming power. This parametrization reduces  the search-space for $x$ from $\R^M$ to $\R^2$. Therefore, the resulting convex program for each $\jammerStrategy\in\jammerStrategyOptionSet$ can be solved efficiently.
	
\begin{figure}
		\centering
		\includegraphics[scale=0.4]{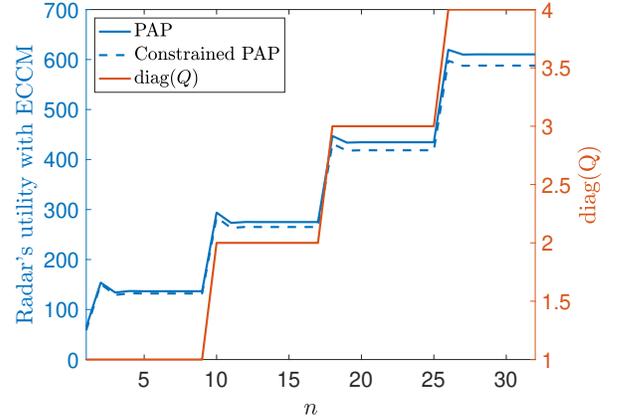}
		\caption{Radar's utility vs. intermediate time index ($n$) for barrage jamming. Simulation parameters are tabulated in Table \ref{tab:simulation_parameters}. Additionally, the target maneuver is updated as $Q_t=\timeIndexThree Q_0$ in the slow timescale. For each maneuver in the slow timescale, simulation was run for a horizon length=8 in the intermediate timescale. Using Theorem~\ref{thm:monotone} to parametrize the radar's ECCM strategy yields a constrained solution within the class of affine ECCM strategies at a low computation cost.}
		\label{fig:radar_utility_approximate}
	\end{figure}

	We again simulate our model for a horizon length of $4$ on the slow timescale with $Q_{\timeIndexThree}=\timeIndexThree \, I_{6\times 6}$. One unit of time in the slow timescale is chosen to be 8 units in the intermediate timescale. This was to allow sufficient time for the transients in the intermediate timescale to settle down. 
	Fig.~\ref{fig:radar_utility_approximate} plots the radar's utility~\eqref{eq:radar-utility-function} vs. $n$ for the radar's ECCM problem~\eqrefPAscSIMPLIFIED\ and the constrained solution~\eqref{eq:parametrize_soln} within the class of affine ECCM strategies. 
    
To summarize, we illustrated an application of Theorem~\ref{thm:monotone} to compute a constrained solution for  ECCM. From Fig.~\ref{fig:radar_utility_approximate}, we observe that the loss in accuracy resulting from a constrained solution is small for small values of the state noise covariance $Q$. Therefore, for targets with small state noise covariance $Q$, the radar can implement the parametrized radar's ECCM strategy without significant degradation in its utility.

\subsection{ECCM strategy for barrage jamming (Sec.\ref{sec:fast_timescale}) when the jammer has imperfect information about the radar's channel}
	\label{sec:numerical-result-noisy-channel-estimate}
	So far, our ECCM formulation assumes that the jammer knows the radar's channel model $\condPMF{\radarObservation}{\jammerStrategy}$. This assumption allowed us to model the radar's ECCM problem as a PAP and derive structural results for the ECCM strategy.
	 We now consider the case where the jammer has imperfect information about the radar's channel model, denoted as  $\condPMFhat{\radarObservation}{\jammerStrategy}$. Given imperfect information of the radar's channel, clearly, the jammer's ECM actions are less effective. But there is also an interesting secondary effect: {\em if  the radar does not know the jammer's estimate of the radar's channel, then the radar's ECCM strategy also becomes less effective.}
	Below we illustrate this via numerical examples.
	
We denote the jammer's imperfect model of the radar's channel as $\condPMFhat{\radarObservation}{\jammerStrategy}$ where
	\begin{align}
	\label{eq:noisy-channel-model}
	    \begin{aligned}
	    &\hat{\probMatrix}=\probMatrix+\Delta\\
    	&\Delta = \begin{bmatrix} 
    	        -0.1099  &  0.0361  &  0.0429 &   0.0310\\
               -0.0079  &  0.0588 &  -0.0165 &  -0.0344\\
                0.0192  &  0.0428 &  -0.1213 &   0.0593\\
                0.0973  &  0.0521 &  -0.0882 &  -0.0612
        \end{bmatrix}
    	\end{aligned}
	\end{align}
    Recall the stochastic matrix $\probMatrix$ denotes  the radar's channel model $\condPMF{\radarObservation}{\jammerStrategy}$ as defined in~\eqref{eq:stochastic-matrix}.
	$\Delta$ models the error in the jammer's estimate of the radar's channel model. To ensure that $ \hat{\probMatrix}$ is a valid  stochastic matrix, the elements of $\Delta$ are chosen so that $\hat{\probMatrix}\geq 0$ (element-wise) and $\hat{\probMatrix}\mathbb{1}=\mathbb{1}$. Here, $\mathbb{1}$ denotes a column vector of 1s of size $M$.
	
	We consider two scenarios:
	
\subsubsection{Scenario 1: Jammer knows $\hat{\probMatrix}$, Radar knows $\probMatrix$ but does not
know  $\hat{\probMatrix}$}
In this scenario,  the radar does not know the jammer's estimate of the radar's channel. Since both jammer and radar are operating with misspecified information, it is intuitive that both the jammer's ECM and radar's ECCM strategy are less effective.
Due to the information mismatch, the mis-specified radar's ECCM problem is:
	\begin{equation}
	   \label{eq:PP}
	   \begin{split}
	        &
	        \max_{\substack{\jammerStrategy\in\jammerStrategyOptionSet,\\\pi\in\R^M}}\; \radarUtility_{\Prob}(\radarStrategy,\jammerStrategy)
	         \\ &
        	\text{s.t.}\;\arg\max_{\bar{J}\in\jammerStrategyOptionSet}\,\jammerUtility_{\Prob}(\radarStrategy,\bar{\jammerStrategy})=\jammerStrategy
        	\end{split}
	    \end{equation}
	    
The subscript $\Prob$ in~\eqref{eq:PP} denotes the probability measure w.r.t.   which the utility functions~\eqref{eq:utility-functions} are evaluated. We denote the solution of~\eqref{eq:PP} as $(\radarStrategy^*_{\Prob{\Prob}},\jammerStrategy^*_{\Prob{\Prob}})$. 

\subsubsection{Scenario 2: Jammer knows $\hat{\probMatrix}$, Radar knows $\probMatrix$ and   $\hat{\probMatrix}$}
In this scenario,  the radar knows the jammer's imperfect estimate of the radar's channel. So it is intuitive that the radar can exploit this additional information to improve its ECCM while the jammer is less effective since it is  operating with mis-specified information. The ECCM problem is given by: 

\begin{equation}
    \label{eq:PPhat}
    \begin{split}
        &\max_{\substack{\jammerStrategy\in\jammerStrategyOptionSet,\\\pi\in\R^M}}\; \radarUtility_{\Prob}(\radarStrategy,\jammerStrategy)\\
    	&\text{s.t.}\;\arg\max_{\bar{J}\in\jammerStrategyOptionSet}\,\jammerUtility_{\hat{\Prob}}(\radarStrategy,\bar{\jammerStrategy})=\jammerStrategy
    \end{split}
\end{equation}
The subscripts $\Prob$, $\hat{\Prob}$ denote the probability measure w.r.t.  which the utility functions~\eqref{eq:utility-functions} are evaluated. We denote the solution of~\eqref{eq:PPhat} as $(\radarStrategy^*_{\Prob\hat{\Prob}},\jammerStrategy^*_{\Prob\hat{\Prob}})$, respectively. 

\subsubsection{Numerical results comparing Scenarios 1 and 2}
\begin{figure}[ht]
		\centering
		\includegraphics[scale=0.4]{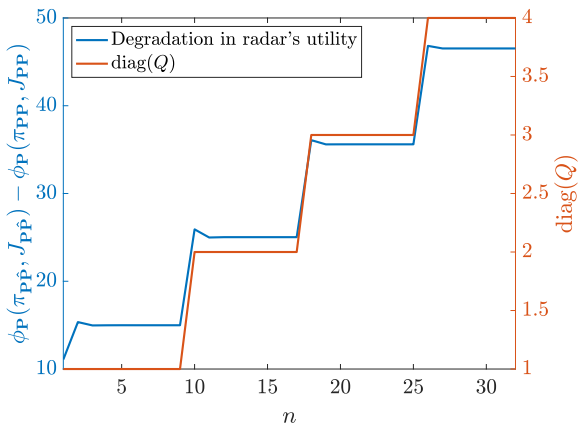}
		\caption{Degradation in radar's utility vs. intermediate time index ($n$) for barrage jamming. Simulation parameters are tabulated in Table \ref{tab:simulation_parameters}. Additionally, target maneuver is updated as $Q_t=\timeIndexThree Q_0$ in the slow timescale. For each maneuver in the slow timescale, simulation was run for a horizon length=8 in the intermediate timescale. The jammer's noisy estimate of the channel model is given by~\eqref{eq:noisy-channel-model}. When the radar does not know $\hat{\Prob}$, its utility decreases. Also, degradation in radar's utility increases with increase in state noise covariance $Q$ of the target. 
		.}
		\label{fig:radar-utility-degrade}
	\end{figure}

\begin{figure}[ht]
		\centering
		\includegraphics[scale=0.4]{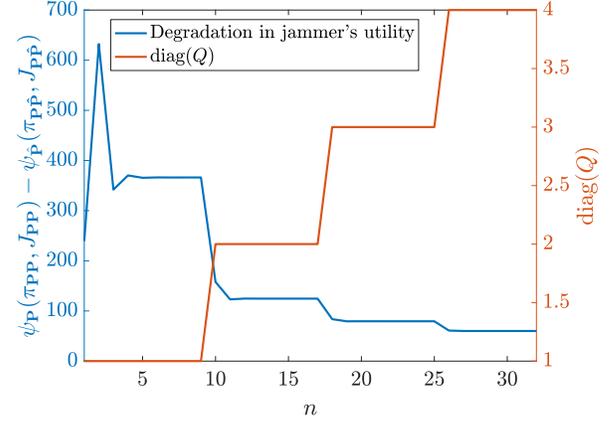}
		\caption{Degradation in jammer's utility vs. time ($n$) for barrage jamming. The  parameters are specified  in Table \ref{tab:simulation_parameters}. The target maneuver (state covariance matrix)  is updated as $Q_t=\timeIndexThree Q_0$ on the slow timescale. For each maneuver in the slow timescale, simulation was run for a horizon length=8 in the intermediate timescale ($n$). The jammer's noisy estimate of the channel model is given by~\eqref{eq:noisy-channel-model}. When the jammer does not know $\Prob$, the jammer's utility decreases. Also, degradation in jammer's utility decreases with an increase in state noise covariance $Q$. 
}
		\label{fig:jammer-utility-degrade}
	\end{figure}

We begin with the study of performance degradation of the radar's ECCM strategy
in Scenario~1. Fig.~\ref{fig:radar-utility-degrade} plots the difference of the radar's utility obtained from~\eqref{eq:PPhat} and~\eqref{eq:PP}, i.e., $\radarUtility_{\Prob}(\radarStrategy^*_{\Prob\hat{\Prob}},\jammerStrategy^*_{\Prob\hat{\Prob}})-\radarUtility_{\Prob}(\radarStrategy^*_{\Prob{\Prob}},\jammerStrategy^*_{\Prob{\Prob}})$. As expected, poor information decreases the utility of the radar. Also, degradation increases with increase in state noise covariance $Q$ of the target~\eqref{eq:target_model} . This is because radar's utility~\eqref{eq:radar-utility-function} is an increasing function of $\maxEV{\Variance}$.
Next, we study the performance degradation of the jammer in Scenario 2. Fig.~\ref{fig:jammer-utility-degrade} displays 
$\jammerUtility_{\Prob}(\radarStrategy^*_{\Prob{\Prob}},\jammerStrategy^*_{\Prob{\Prob}})-\jammerUtility_{\hat{\Prob}}(\radarStrategy^*_{\Prob\hat{\Prob}},\jammerStrategy^*_{\Prob\hat{\Prob}})$. As expected, poor information decreases the utility of the jammer. Also, degradation in jammer's utility decreases with increase in state noise covariance $Q$ of the target~\eqref{eq:target_model}. This is because the jammer's utility~\eqref{eq:jammer-utility-function} is a decreasing function of $\maxEV{\Variance}$.
	
	To summarize, Fig.~\ref{fig:radar-utility-degrade} and  Fig.~\ref{fig:jammer-utility-degrade} show that the optimal radar's ECCM and jammer's ECM  are less effective when the radar and the jammer have imperfect information about the other. Also, since the utility functions~\eqref{eq:utility-functions} depend on  $\maxEV{\Variance}$, we observe that degradation in the radar's utility increases with an increase in state noise covariance $Q$. For the jammer, degradation in the jammer's utility decreases with an increase in state noise covariance $Q$.

    \subsection{ECCM  for deception jamming (Sec.\ref{sec:example-2}) using  PAP}
    \label{sec:sim:deception-jamming-radar-utility}
\begin{figure}[h]
        \centering
        \includegraphics[scale=0.4]{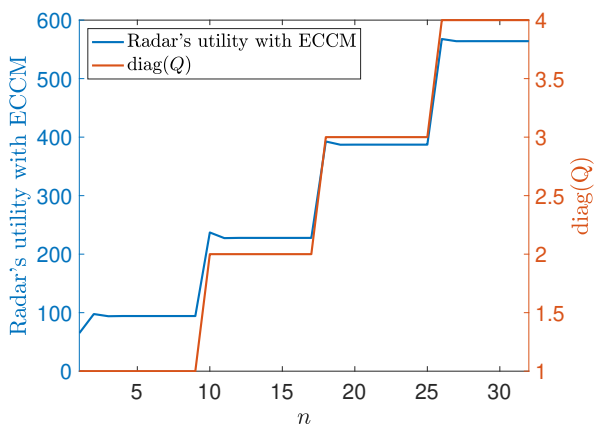}
        \caption{Radars's utility vs. intermediate time index $(n)$ for deception jamming. Simulation parameters are tabulated in Table~\ref{tab:simulation_parameters_updated}. Additionally, the target maneuver is updated as $Q_t=tQ_0$ in the slow timescale. For each maneuver in the slow timescale, simulation was run for a horizon length=$8$ in the intermediate timescale. For our choice of model parameters, deception jamming lowers the radar's utility in comparison to barrage jamming. Hence, deception jamming is more effective as an ECM than barrage jamming.}
        \label{fig:ur-example2}
    \end{figure}

    We now simulate our second example from Sec.\ref{sec:example-2}: ECCM against deception jamming by an adversarial jammer. The simulation parameters are tabulated in Table~\ref{tab:simulation_parameters_updated}. As much of the model remains unchanged, one can reproduce similar results for ECCM against deception jamming. For brevity, we only present a plot of the radar's utility function for a comparative study between the two types of jamming. Fig.~\ref{fig:ur-example2} displays the radar's utility with ECCM vs. intermediate time index $(n)$ for the case of deception jamming. For the simulation parameters specified in Table~\ref{tab:simulation_parameters} and Table~\ref{tab:simulation_parameters_updated}, a comparison between Fig.~\ref{fig:radar_utility_approximate} and Fig.~\ref{fig:ur-example2} shows that deception jamming leads to reduction in the utility of the radar. Hence, for our choice of model parameters, deception jamming is advantageous for the adversarial jammer.
    
    \begin{table}[]
        \centering
        \caption{Simulation Parameters for the Radar's ECCM Problem against Deception Jamming.}
		\begin{tabular}{r|r|l}
			\hline
			\textbf{Parameters} & \textbf{Eq.} & \textbf{Value}\\
			\hline
			$T$ &~\eqref{eq:target_model_updated}& 1\\
			$B_1,B_2,Q_0$ &~\eqref{eq:target_model_updated}& $I_{6\times 6}$\\
			$\begin{bmatrix}C_1 & C_2\end{bmatrix}$ &~\eqref{eq:target_model_updated} & $\begin{bmatrix}I_{6\times 6}& 0.5\,I_{6\times 6}\end{bmatrix}$\\
			$\{\jammerStrategyOption_1,\ldots,\jammerStrategyOption_4\}$ &~\eqref{eq:jammer-strategy-set} & $\{1,2,3,4\}$\\
			$c_1$ &~\eqref{eq:pa-sc-simplified-objective} & $1\times10^{2}$\\
			$c_2$ &~\eqref{eq:pa-sc-simplified-ic} & $1\times10^{4}$\\
			$\probMatrix$ &~\eqref{eq:stochastic-matrix} & $\begin{bmatrix}
			0.3878 & 0.3215 & 0.1858 & 0.1049\\
			0.2980 & 0.3617 & 0.2146 & 0.1256\\
			0.2040 & 0.2583 & 0.3307 & 0.2070\\
			0.1029 & 0.1408 & 0.2140 & 0.5422
			\end{bmatrix}$\\
			\hline
		\end{tabular}
		\label{tab:simulation_parameters_updated}
    \end{table}
    
	\subsubsection*{Summary of Numerical Results}
	\begin{compactenum}
	    \item In Sec.\ref{sec:model_setup}, we explored numerically how the  expected signal-to-noise ratio $\SNRn$~\eqref{eq:expected_SNR} varies with increasing uncertainty in the target maneuver. As the state covariance $Q$ increases, the radar's marginal reward~\eqref{eq:radar-utility-function} increases, whereas the marginal cost to increase the effort remains the same. For the jammer, an increase in state covariance $Q$ decreases the marginal reward~\eqref{eq:jammer-utility-function} whereas the marginal cost to increase the effort remains the same. Hence, overall the SNR of the radar increases with an increase in the state noise covariance $Q$.
	    \item In Sec.\ref{sec:rl-eccm}, we parametrized the radar's ECCM strategy within the class of increasing affine functions~\eqref{eq:parametrize_soln}. This reduced the search space for the radar's ECCM strategy from $\R^M$ to $\R^2$. Through simulation results, we observed that at a low state noise covariance of the target, the radar can implement a parametrized ECCM strategy without significant degradation in its utility. 
	    The parametrized solution also has the advantage of being computationally efficient. 
	   \item In Sec.\ref{sec:numerical-result-noisy-channel-estimate}, we investigated via numerical examples what happens when the radar and the jammer have imperfect information about the other. Numerical results showed that the optimal ECCM and ECM of the radar and the jammer are less effective when the radar and the jammer have imperfect information about the other. Due to parametrization of the utility functions \eqref{eq:utility-functions}, we observed that degradation in radar's utility increases with an increase in state noise covariance $Q$; whereas degradation in the jammer's utility decreases with an increase in state noise covariance $Q$.
	   \item In Sec.\ref{sec:sim:deception-jamming-radar-utility}, we computed the utility of the radar when it employs ECCM to mitigate deception jamming (ECM) by an adversarial radar. For our choice of model parameters, we observed that the deception jamming was more effective as an ECM than the barrage jamming.
    \end{compactenum}
	\section{Conclusion and Extensions}	
	We proposed the principal agent problem (PAP) as a principled approach for the radar's electronic counter-countermeasure (ECCM) problem.
	In the PAP studied in microeconomics, the principal designs a contract to induce a specific action from an agent when the agent's action is observed in noise. In complete analogy, in this paper, 
the radar's ECCM strategy (obtained by solving the PAP)  mitigates the effect of the jammer's electronic countermeasure (ECM) by inducing specific actions by the jammer. 
	By incorporating performance and measurement cost in the utility function, we formulated a PAP with the radar-jammer as a principal-agent pair. Using a utility-maximization approach, we modeled the trade-off between performance and associated cost. Our main results were the following: i) we formulated the ECCM problem as a convex optimization problem, and ii) we derived conditions under which the optimal ECCM strategy is an increasing function of the jamming power observed by the radar. Finally, we simulated the PAP to compute the optimal radar's ECCM strategy. The importance of structural results in parametrizing the solution of the PAP was dealt with qualitatively using simulations. Towards the end, we also simulated a radar's ECCM problem wherein the radar and the jammer have mismatched information. 
	
	The PAP approach to  ECCM in this paper
	 can be extended to more general settings involving multiple networked radars and jammers in a shared environment \cite{2010:AA-et-al}. The presence of multiple radars in a shared environment creates an additional issue of inter-radar interference. This is a challenging problem for future work. 
	
	{\bf Acknowledgment}.
	This research was supported in part by the U.S.\ Army Research Office grant  
	W911NF-21-1-0093 and National Science Foundation grant CCF-2112457.
	
	\appendices
	\section{Proof of Theorem~\ref{thm:convex} in Sec.\ref{sec:intermediate_timescale}}
	The change of variable $x_i=\log(\radarStrategy_i)$ in~\eqrefPAgeneral\ for the choice of utility functions in \eqUTILITYfunctions\
	yields  the optimization problem~\eqrefPAscSIMPLIFIED.
	The objective~\eqref{eq:pa-sc-simplified-objective} is concave in unknowns for a fixed $\jammerStrategy$. The constraint~\eqref{eq:pa-sc-simplified-ic} can be re-written as $M-1$ affine inequality constraints. Hence, the resulting optimization problem is a convex program.
	
	\section{Proof of Lemma~\ref{lemma:jammer-utility-concave} in Sec.\ref{sec:structural-results}}
	Define $x_0=0$.
	\begin{align*}
	\mathbb{E}_{\radarObservation|\jammerStrategy}\left[\operatorname{e}_{\boldRadarObservation}^\prime x\right] &= \sum_{m=1}^{M} \condPMF{\jammerStrategyOption_m}{\jammerStrategy} \sum_{j=1}^{m}\left(x_{j}-x_{j-1}\right)\\
	&=\sum_{j=1}^{M}\left(x_{j}-x_{j-1}\right) \sum_{m=j}^{M} \condPMF{\jammerStrategyOption_m}{\jammerStrategy}\\
	&=x_{1}+\sum_{m=2}^{M}\left(x_{m}-x_{m-1}\right) \Pi_{m}\left(\jammerStrategy\right)\\
	\therefore \jammerUtility(\radarStrategy,\jammerStrategy)&:=\mathbb{E}_{\radarObservation|\jammerStrategy}\left[\frac{c_2}{\maxEV{\Variance}}(- x(\boldRadarObservation)+\log(\boldRadarObservation))\right]-\jammerStrategy^{2}\\
	&\nonumber=\frac{c_2}{\maxEV{\Variance}}\left(-x_1- \sum_{m=2}^{M}\left(x_{m}-x_{m-1}\right) \Pi_{m}\left(\jammerStrategy\right)\right.\\
	&\quad \Bigg.+\mathbb{E}_{\radarObservation|\jammerStrategy}\left[\log(\boldRadarObservation)\right]\Bigg)-\jammerStrategy^{2}
	\end{align*} 
Note that $\jammerUtility(\radarStrategy,\jammerStrategy)$ in~\eqref{eq:jammer-utility-function} is the sum of functions which are concave in $\jammerStrategy$ and some constants. Hence, $\jammerUtility(\radarStrategy,\jammerStrategy)$ is a concave in $\jammerStrategy$.
	
	\section{Proof of Theorem~\ref{thm:monotone} in Sec.\ref{sec:structural-results}}
	To prove Theorem~\ref{thm:monotone}, we first rewrite~\eqref{eq:pa-sc-simplified-ic},~\eqref{eq:relaxed-pap} as a set of affine inequality constraints:
	\begin{align}
	    \intertext{Constraint~\eqref{eq:pa-sc-simplified-ic} for the  ECCM problem is equivalent to:}
	    &\nonumber\mathbb{E}_{\radarObservation|\jammerStrategy}\bracketSquare{\frac{c_2}{\maxEV{\Variance}}\,(\log(\boldRadarObservation)- \operatorname{e}_{\boldRadarObservation}^\prime x)}-\jammerStrategy^{2}\geq\\
	    &\nonumber\qquad\qquad\mathbb{E}_{\radarObservation|\bar{\jammerStrategy}}\bracketSquare{\frac{c_2}{\maxEV{\Variance}}\,(\log(\boldRadarObservation)- \operatorname{e}_{\boldRadarObservation}^\prime x))}-\bar{\jammerStrategy}^{2},\;\forall \bar{\jammerStrategy}\neq\jammerStrategy
	    \intertext{Constraint~\eqref{eq:relaxed-pap} for the relaxed  ECCM problem is}
	    &\nonumber\mathbb{E}_{\radarObservation|\jammerStrategy}\bracketSquare{\frac{c_2}{\maxEV{\Variance}}\,(\log(\boldRadarObservation)- \operatorname{e}_{\boldRadarObservation}^\prime x)}-\jammerStrategy^{2}\geq\\
	    &\nonumber\qquad\qquad\mathbb{E}_{\radarObservation|\bar{\jammerStrategy}}\bracketSquare{\frac{c_2}{\maxEV{\Variance}}\,(\log(\boldRadarObservation)- \operatorname{e}_{\boldRadarObservation}^\prime x))}-\bar{\jammerStrategy}^{2},\;\forall \bar{\jammerStrategy}>\jammerStrategy
	\end{align}
	The first-order necessary conditions for optimization problem \cite{2004:SB-LV} helps us to derive the main result of this section. With $\mu_{{\textstyle\mathstrut}\bar{\jammerStrategy}}\geq 0$ as Lagrange multipliers and $\scrL$ to denote the Lagrangian of the convex optimization problem~\eqrefPAscSIMPLIFIED, first-order necessary condition for optimality is given by~\eqref{eq:KKT-first-order}:
	\begin{align}
	\label{eq:KKT-first-order}
	&\frac{\partial \scrL}{\partial x_{m}} = 0\\
	&\nonumber\Rightarrow \condPMF{\jammerStrategyOption_m}{\jammerStrategy}\left(2\e^{2x_m}-c_1\maxEV{\Variance}\right)\\
	\nonumber&=\frac{c_2}{\maxEV{\Variance}}\sum_{\bar{J}}\mu_{{\textstyle\mathstrut}\bar{\jammerStrategy}}\bracketSquare{\condPMF{\jammerStrategyOption_m}{\bar{\jammerStrategy}}-\condPMF{\jammerStrategyOption_m}{\jammerStrategy}}\\
	\label{eq:kkt}
	&\nonumber\Rightarrow \e^{2x_m} = \frac{1}{2}\Bigg[c_1\maxEV{\Variance}\Bigg.\\
	&\qquad\qquad\quad\Bigg.+\sum_{\bar{\jammerStrategy}} \bracketRound{\frac{\condPMF{\jammerStrategyOption_m}{\bar{\jammerStrategy}}}{\condPMF{\jammerStrategyOption_m}{\jammerStrategy}}-1} \frac{\mu_{{\textstyle\mathstrut}\bar{\jammerStrategy}}\,c_2}{\maxEV{\Variance}}\Bigg]\\
	\label{eq:kkt-difference}
	&\nonumber\Rightarrow \e^{2x_m}-\e^{2x_l}=\sum_{\bar{\jammerStrategy}}\Bigg[\frac{\condPMF{\jammerStrategyOption_m}{\bar{\jammerStrategy}}}{\condPMF{\jammerStrategyOption_m}{\jammerStrategy}}\Bigg.\\
	&\qquad\qquad\qquad\qquad\qquad\Bigg.-\frac{\condPMF{\jammerStrategyOption_m}{\bar{\jammerStrategy}}}{\condPMF{\jammerStrategyOption_l}{\jammerStrategy}}\Bigg] \frac{\mu_{{\textstyle\mathstrut}\bar{\jammerStrategy}}\,c_2}{2\,\maxEV{\Variance}}
	\end{align}
	We exploit~\eqref{eq:kkt-difference} to prove Theorem~\ref{thm:monotone}. Let $\jammerStrategy$ be the jamming power to be incentivized. To prove first part, we use the same steps as ~\eqref{eq:kkt}-\eqref{eq:kkt-difference} for the relaxed radar's ECCM problem~\eqref{eq:relaxed-pap} to obtain:
	\begin{align}
	&\nonumber\e^{2x_m}-\e^{2x_l}=\sum_{\bar{\jammerStrategy}>\jammerStrategy}\Bigg[\frac{\condPMF{\jammerStrategyOption_m}{\bar{\jammerStrategy}}}{\condPMF{\jammerStrategyOption_m}{\jammerStrategy}}\Bigg.\\
	&\nonumber\qquad\qquad\Bigg.-\frac{\condPMF{\jammerStrategyOption_m}{\bar{\jammerStrategy}}}{\condPMF{\jammerStrategyOption_l}{\jammerStrategy}}\Bigg]\frac{\mu_{{\textstyle\mathstrut}\bar{\jammerStrategy}}\,c_2}{2\,\maxEV{\Variance}}\geq 0,\;\forall m>l
	\end{align}
	Therefore, $x_m$ is non-decreasing in $m$ i.e. $x_1\leq x_2 \leq \ldots \leq x_M$, which implies $\jammerUtility(\radarStrategy,\jammerStrategy)$ is concave in $\jammerStrategy$ for the relaxed radar's ECCM problem using Lemma~\ref{lemma:jammer-utility-concave}.
	
	To prove second assertion of Theorem~\ref{thm:monotone}, note that for $\jammerStrategy=\jammerStrategyOption_1$,~\eqref{eq:relaxed-pap} is same as PAP~\eqrefPAscSIMPLIFIED. For $\jammerStrategy\geq \jammerStrategyOption_2$, we know that the any solution of the relaxed radar's ECCM problem should be non-constant or else $\jammerStrategy_1$ becomes the optimal jamming power. This implies at least one of the incentive constraint for the relaxed radar's ECCM problem is binding. Let the corresponding constraint be for $\bar{\jammerStrategy}=\delta>\jammerStrategy$. We have
	\begin{align}
	\nonumber\jammerUtility(\radarStrategy,\delta)&=\jammerUtility(\radarStrategy,\jammerStrategy)\geq \jammerUtility(\radarStrategy,\bar{\jammerStrategy}),\;\forall \bar{\jammerStrategy}>\jammerStrategy\\
	\intertext{Due to concavity of $\jammerUtility(\radarStrategy,\jammerStrategy)$, we also get}
	\nonumber\jammerUtility(\radarStrategy,\jammerStrategy)&\geq \jammerUtility(\radarStrategy,\bar{\jammerStrategy}),\;\forall \bar{\jammerStrategy}<\jammerStrategy
	\intertext{because if $\jammerUtility(\radarStrategy,\jammerStrategy)< \jammerUtility(\radarStrategy,\tau)$ for some $\tau<\jammerStrategy <\bar{\jammerStrategy}$ implies:} 
	\nonumber\jammerUtility(\radarStrategy,\jammerStrategy)&<0.5\;\jammerUtility(\radarStrategy,\delta)+0.5\;\jammerUtility(\radarStrategy,\tau)
	\end{align}
	which contradicts the fact that $\jammerUtility(\radarStrategy,\jammerStrategy)$ is a concave function in $\jammerStrategy$. Hence, the radar's ECCM strategy for the relaxed radar's ECCM problem~\eqref{eq:relaxed-pap} is also optimal for the radar's ECCM problem~\eqrefPAscSIMPLIFIED.
	
	\ifCLASSOPTIONcaptionsoff
	\newpage
	\fi
	
	\bibliography{anurag}{}
	\bibliographystyle{unsrt}
\end{document}